\documentstyle[12pt]{article}
\begin{document}
\pagestyle{empty}
\noindent
{\bf  On Deusons or Deuteronlike Meson-Meson Bound States\footnote
{Invited talk at the Hadron93 International Conf. on Hadron Spectroscopy,
Como, Italy 22.-25.6. 1993. HU-SEFT R 1993-13a}
}
\vskip 0.4cm

\noindent NILS A. T\"ORNQVIST
\vskip 0.3cm

\noindent {\small \it Research Institute for High Energy Physics,
University of Helsinki -
PB 9, FIN-00140 Helsinki, Finland}
\vskip 0.5cm

{\small \noindent {\bf Abstract.}  The systematics of
deuteronlike two-meson bound states, {\it deusons}, is discussed.
Previous arguments that many of the present non-$q\bar q$ states are
such states are elaborated including,
in particular, the tensor potential.
For pseudoscalar  states the important observation is made
that the centrifugal barrier from the P-wave  can be
overcome by the $1/r^2$ and $1/r^3$ terms of the tensor potential.
In the heavy meson  sector one-pion exchange alone
is strong enough to form at least deuteron-like
$B\bar B^*$ and  $B^*\bar B^*$ composites bound by approximately 50 MeV,
while $D\bar D^*$ and $D^*\bar D^*$ states are expected near the threshold.
}
\vskip 0.5cm
 Recently I suggested \cite{NAT1}
that many of the best established light non-$q\bar q$
candidates \cite{PDG}
are in fact deuteronlike meson-meson bound states or {\it deusons}.
The idea, that deuteronlike
bound states of two mesons might exist
is certainly not new, but has been discussed
generally only in passing whithin general
phenomenological models for meson-meson bound
states (See \cite{Okun}--\cite{ISGUR}),
where pion exchange
is not given special attention. After my first
letter, Ericson and Karl \cite{Karl} has also studied the
strength of pion exchange with similar conclusions,
 and Manohar and Weise \cite{Mano} have studied
flavour exotic two $B^*$-meson bound states. The heavy meson systems are
an interesting testing ground for these ideas,
since the predictions are less ambiguous than for light states.

One can write the one-pion exchange potential in an universal way
by collecting all the constants
into an overall number $\gamma$, the "relative coupling number" which is a
measure of the overall potential strength.
Thus for $NN$ one has $\gamma^{NN}_{SI}=
-\frac{25}{9}(\tau_1\cdot \tau_2)(\sigma_1\cdot\sigma_2)$,
for $(D^*\bar D)_\pm ,\ \gamma^{PV}_{I\pm}=\mp \tau_1\cdot \tau_2 $
and for $D^*\bar D^*\ \gamma^{VV}_{SI}=-(\tau_1\cdot
\tau_2)(\Sigma_1\cdot\Sigma_2)$. This number $\gamma$ measures
 the relative strength of the potential
compared to the contribution from one pair of
quarks in a spin triplet and isospin
triplet state for which $\gamma^{qq}_{SI}=-1$. For example for the
deuteron
 $\gamma^{NN}_{10}=25/3$ and for  $D^*\bar D^*$  in I=0, S=0,
$\gamma^{VV}_{00}=6$.
The larger $\gamma$ is, the stronger is the attraction,
and if it is negative there is repulsion. The universal one-pion exchange
potential in $r$-space can then be written compactly:
\begin{eqnarray}
V_{\pi}(r) = -\gamma V_0\ [D\cdot C(r)+S_{12}(\hat r)\cdot T(r)] \ \label{Vr} ,
\end{eqnarray}
where $D$ is a diagonal matrix, $\hat r$ is the unit vector, and the $r$
dependence
is given by the functions
\begin{eqnarray}
C(r)&=& \frac {\mu^2}{m_\pi^2} \frac {e^{-\mu r}}{m_\pi r} \ ,\label{Cr} \\
T(r)&=& C(r) [1+\frac 3{\mu r} +\frac 3 {(\mu r)^2}] \ ,\label{Tr}
\end{eqnarray}
\noindent
and $S_{12}(r)$ is the tensor operator in $r$ space, which, in general,
connects different partial waves.
In Eq.~(\ref{Vr}) we  introduced the constant
$ V_0=m_\pi^3 g^2 / (12\pi f^2) \approx 1.3$ MeV,
the numerical value of which is fixed by the $\pi N$ coupling
constant.

Because of the singular behaviour of the tensor potential it
 must be regularized at small distances. The perhaps most natural
 method is to introduce a form  factor at each $\pi N$ vertex, such as
$(\Lambda^2-\mu^2)/(\Lambda^2+t)$, which in $r$-space can be looked
upon as a spherical pion source with rms radius $R=\sqrt {10}/\Lambda $.

{\bf The deuteron.}
Our prime reference state is of course the deuteron, the existence of which
nobody doubts. It has been studied in great detail over the years
(See Ref.~\cite{Glen} and the recent reviews \cite{Rosa}, \cite {Torleif}).
There one knows that the dominant binding energy comes from
pion exchange between two colourless $qqq$ clusters - a proton and a
neutron.

One defines conventional basis vectors $|^3S_1>$ and $|^3D_1>$
 such that the wave function is in general
$u(r)|^3S_1>+w(r)|^3D_1>$. The  deuteron potential $V_d(r)$ can then be written
in matrix form as:
\begin{eqnarray}
V_{d}(r) = -\frac{25}{3} V_0  \left[
\left( \begin{array}{cc} 1&   0      \\    0     & 1\end{array}\right) C(r) +
\left( \begin{array}{cc} 0& \sqrt 8 \\ \sqrt 8 & -2\end{array}\right) T(r)
\right]\ ,   \label{Vdeut}
\end{eqnarray}
The overall strength of the potential  is given by $\gamma V_0=11.4$ MeV.
There are a  few important
points to be learned from the deuteron and the $NN$ system,
which are essential also for the present application to meson-meson states:
\begin{itemize}
\item The central part of the one pion
potential $\gamma_{10}^{NN} V_0C(r)$ is insufficient to bind the
deuteron.  The overall strength is too small by a large factor of 3.

\item The tensor potential is very important
in providing the binding \cite{Torleif}.
It is much larger in magnitude than
the central part at small distances.

\item  The potential must be regularized at small $r$, by some
cut off procedure. The binding energy is very sensitive to this procedure,
as well as to any added short range potential from $2\pi , \rho, \omega$
etc. exchange, But, once the binding energy is right all
the other static deuteron properties follow correctly.
\end{itemize}

Unfortunately because of this last fact one will
not, in general, be able to predict reliably the binding energies for deusons.
Certainly, if the potential is strong enough one can be quite confident that
such bound states must exist, but their exact binding energy will always
depend on details of regularization, and  on the short range potential
from heavier exchanges. Again, in borderline
cases where the expected binding is small, like for the deuteron, one
similarily cannot be sure that  such bound states actually exist.

For the numerical solutions we use a method (See Refs.~\cite{Basd},\cite{Rich})
where one  discretizes the $r$ dependence to a finite dimensional vector,
wherby the Hamiltonian becomes a finite matrix,
which can be solved by using efficient standard matrix routines
for finding eigenvalues and eigenvectors. The method is particularily
accurate for finding the ground state in problems with coupled channels,
which is precisely what is needed here.
The method is comphrehensively discussed
for the one-channel problem in Ref.~\cite{Rich}.

{\bf  PV deusons.}
Since parity forbids two pseudoscalars to be bound by one-pion exchange,
the lightest deusons are pseudoscalar-vector states.
Again in  such $PV$ systems the  pion is too light to
be a constituent, because the small reduced mass of a $\pi V$ system would
give a too large kinetic term, which cannot be overcome by the potential.
Thus the deuson spectrum can start at the earliest with $K\bar K^*$ or
at $\approx 1400$ MeV.
In general, the relative coupling number,
 $\gamma$, is 3 times larger for I=0 states
than for I=1, and the $J^{PC}$ quantum numbers of interest for $PV$ systems
are $0^{-+}$ and $1^{++}$. For these $\gamma=3$ and the potentials are:
\begin{eqnarray}
V_{0^{-+}}& =& -3 V_0 [C(r)+2T(r)]=
-3V_0 \frac {\mu^2}{m_\pi^2}\frac{e^{-\mu r}}{m_\pi r} \left[ 3
 +\frac {6}{\mu r}+\frac {6}{(\mu r)^2} \right]\ , \label{Vpv0} \\
V_{1^{++}}& =& -3V_0  \left[
\left( \begin{array}{cc} 1&   0      \\    0     & 1\end{array}\right) C(r) +
\left( \begin{array}{cc} 0& -\sqrt 2 \\ -\sqrt 2 & 1\end{array}\right) T(r)
\right]\ .  \label{Vpvax}
\end{eqnarray}
\noindent
Here $\mu$ is not exactly the pion mass, since the vector and
pseudoscalar do not have the same mass. This leads to a recoil effect
through which one has instead:
$\mu^2=m_\pi^2-(M_V-M_P)^2 $.
For the axial vector states $1^{++}$ (\ref{Vpvax}) there are two channels
like for the deuteron, and
we define  basis vectors  such that in general
$|1^+>=u(r) |^3S_1>+w(r) |^3D_1>$.

The pseudoscalar  channel is a single channel case, Eq.~(\ref{Vpv0}), to
which the tensor part contributes
with same sign as the central part since $<\ ^3P_0|S_{12}|^3P_0>=+2$.
Therefore these add giving a remarkably strong potential (\ref{Vpv0}),
which is attractive for C=+ i.e., $J^{PC}=0^{-+}$.
Thus an especially interesting new situation appears:
The tensor part with its $r^{-2}$
and $r^{-3}$ terms contributes, with
much stronger attraction than the central part, directly to the $^3P_0$
wave, and not through a second order coupling of S and D waves as for
the deuteron.
This means that
the $1/ r^{2}$ and the $1/ r^{3}$  terms of the tensor potential
can compensate, at least partially, the centrifugal barrier $2/(Mr^2)$.

Numerically by solving the Schr\"odinger equation  one finds using
the one-pion potential (\ref{Vpv0}-\ref{Vpvax})
alone, regularized with $\Lambda =1.2$ GeV that
\begin{itemize}
\item that there certainly must exist
an $\eta_b(\approx 10545)$, that there very likely exists an $\eta_c(\approx
3870)$, and that possibly in the iota peak, $\eta(1440)$, there is a
$K\bar K^*$ deuson.
\item that certainly there must exist
a $\chi_{1b}(\approx 10562)$,  that rather likely there exists a
$\chi_{1c}(\approx 3870)$ and that
possibly the $f_1(1420)$ could be a $K\bar K^*$ deuson.
\end{itemize}
{\bf VV deusons.} For composites of two vector mesons the
strongest attraction, as measured by the relative coupling number
$\gamma^{VV}_{SI}$,
is in the spin and isospin singlet channels ($\gamma^{VV}_{00}=+6$),
followed by the spin triplet, isosinglet channels ($\gamma^{VV}_{10}=+3$).
For I=1 one has either repulsion or very weak attraction.
The S-waves appear for $J^{PC}=0^{++},\ 1^{+-}$ and $2^{++}$, but
these can mix with D-waves and the potential for the two first
mentioned channels are:
\begin{eqnarray}
V_{0^{++}}& =& -6 V_0   \left[
\left( \begin{array}{cc} 1&   0      \\ 0 & -\frac 1 2\end{array}\right) C(r) +
\left(\begin{array}{cc}
0&\sqrt\frac 1 2\\ \sqrt\frac 1 2&1\end{array}\right) T(r)
\right]\ , \label{Vvscalar}\\
V_{1^{+-}}& =& -3 V_0   \left[
\left( \begin{array}{cc} 1&   0      \\    0     & 1\end{array}\right) C(r) +
\left( \begin{array}{cc} 0& -\sqrt 2 \\ -\sqrt 2 & 1\end{array}\right) T(r)
\right]\ . \label{Vv1pm}
\end{eqnarray}
\noindent For the single channel cases the most interesting
is the pseudoscalar $^3P_0$:
\begin{eqnarray}
V_{0^{-+}} &=&
- 3 V_0[C(r) + 2T(r)] = -3 V_0\frac {e^{-m_\pi r}}
{fr} [3+\frac 6{m_\pi r}+\frac 6{(m_\pi r)^2}] \ ,\ \label{Vvps}
\end{eqnarray}
\noindent
which is in fact very similar to
that for pseudoscalar $(P\bar V)_+$ states, (Eq.~\ref{Vpv0}),
and is  remarkably strong, as was already discussed for the $P\bar V$ case.
There remains two spin parities which  are of interest,
$1^{--}$ and $2^{++}$, which are discussed in Ref.~\cite{NAT2}, and
in which the tensor potential connects 3 respectively 4 partial waves.
In summary by solving the Schr\"odinger equation one finds
$VV$ deusons of the following kinds:
\begin{itemize}
\item In the pseudoscalar sector an
 $\eta_b(\approx 10590)$ should exist and $\eta_c(\approx 4015)$
should very likely also exist
and an $ \eta (\approx 1790)$ is possibly bound.
These states should all mix to some extent with their $P\bar V$ counterparts,
making the lighter (mostly $P\bar V$) more bound.

\item In the scalar sector there should exist  a
$\chi_{b0} (\approx 10582)$ $B^*\bar B^*$ bound state and a
$\chi_{c0}(\approx 4015)$ also very likely exists.
In the light sector the $f_0(1720)$ (the "theta") could be a $K^*\bar K^*$,
while the $f_0(1520)$ could be a $(\rho\rho -\omega\omega )/\sqrt 2$  deuson.

\item In the axial sector
$h_b(\approx 10608)$ should exist, possibly also a
$h_c(\approx 4015)$, while $ h_1(\approx 1790)$
is less likely.

\item In the tensor sector the numerical calculations show, perhaps
surprisingly, that there should exist a
$\chi_{b2} (\approx 10602)$ $B^*\bar B^*$ bound state, and that
a $\chi_{c2}(\approx 4015)$ at threshold is also
very likely. With some extra
attraction the  $f_2(1720)$ \cite{theta}
could be a $K^*\bar K^*$, and the $f_2(1520)$
could be a $(\rho\rho + \omega\omega )/\sqrt 2 $ deuson.
\end{itemize}
The widths are expected to be quite narrow. This is especially the case
for the four  $D\bar D^*$ and $B\bar B^*$ states, which because of parity
cannot
decay to $D\bar D$, respectively $B\bar B$. Of course through annihilation
of the heavy quarks, decays to light mesons are possible.
However, such annihilation should be
suppressed by form factors, since
these states are much larger in size than normal $Q\bar Q$ states
(assuming the states are weakly bound).
The $D^*\bar D^*$ and $B^*\bar B^*$ deusons can generally decay into
$D\bar D$ and $B\bar B$, which should be their main decay mode giving
widths of a few tens of MeV.

In channels with exotic flavour or CP quantum numbers
pion exchange is generally repulsive or
quite weak. Therefore one does not expect
that such exotic deusons  exist, although
$B^* B^*$ may be an exception.
Neither does the deuson
model predict new non-$q\bar q$  states which should have been
seen. E.g., for I=1 channels
one pion exchange is generally
a factor 3 weaker than for I=0, and one  certainly does not expect such states
within the light meson sector.

Where could the predicted heavy deusons of Table 8 be produced and seen
experimentally? Unfortunately, this will not be easy, but at least
two good places are:  $N\bar N$
in flight at the Fermilab antiproton accumulator, and $\Upsilon$ decay
looking at final states including e.g.
$J/\psi\ \omega $ (for the $D\bar D^*$ deusons)
or $D\bar D$, $D\bar D^*$ (for the $D^*\bar D^*$ deusons).

To find these states would be important, not only because they would
confirm the expectations from pion exchange and constrain the parameters
of the model presented here. More importantly, if these
deusons are found, they at the same time
would give strong support for the interpretation that
many, perhaps all, of the
present light non-$q\bar q$ candidates really are deuteronlike states.
 This would then
imply that experimental evidence for baglike multiquark states and glueballs
would have to be looked for at higher energies.

More details about the results presented here are given in Ref.~\cite{NAT2}.

\eject

\end{document}